\documentclass[a4paper,11pt]{article}
\usepackage{amsfonts}
\usepackage{graphicx}
\usepackage{amsmath}

 \DeclareFontFamily{U}{rsf}{}
\DeclareFontShape{U}{rsf}{m}{n}{
  <5> <6> rsfs5 <7> <8> <9> rsfs7 <10-> rsfs10}{}
\DeclareMathAlphabet\Scr{U}{rsf}{m}{n} \makeatletter
\@addtoreset{equation}{section} \makeatother

\def\be{\begin{equation}}
\def\ee{\end{equation}}
\def\ba{\begin{array}}
\def\ea{\end{array}}

\newcommand{\bea}{\begin{eqnarray}}
\newcommand{\eea}{\end{eqnarray}}

\begin{document}

\begin{titlepage}
 \thispagestyle{empty}
 \begin{flushright}
     \hfill{CERN-PH-TH/2011-308}\\
 \end{flushright}

 \vspace{50pt}

 \begin{center}
     { \huge{\bf      {Black Holes and Groups of Type $E_{7}$}}}

     \vspace{70pt}

     {\Large {\bf Sergio Ferrara$^{a,b}$ and Alessio Marrani$^{a}$}}

     \vspace{30pt}

  {\it ${}^a$ Physics Department, Theory Unit, CERN,\\
     CH -1211, Geneva 23, Switzerland\\\texttt{sergio.ferrara@cern.ch}\\\texttt{alessio.marrani@cern.ch}}

     \vspace{20pt}

   {\it ${}^b$ INFN - Laboratori Nazionali di Frascati,\\
     Via Enrico Fermi 40, I-00044 Frascati, Italy}

     \vspace{15pt}

     \vspace{100pt}

     {ABSTRACT}
 \end{center}

 \vspace{10pt}
\noindent We report some results on the relation between extremal
black holes in locally supersymmetric theories of gravity and groups
of type $E_{7}$, appearing as generalized electric-magnetic duality
symmetries in such theories. Some basics on the covariant approach
to the stratification of the relevant symplectic representation are
reviewed, along with a connection between special K\"{a}hler geometry and a \textit{%
``generalization''} of groups of type $E_{7}$.
 \vfill
\begin{center}{\sl
Contribution to the Proceedings of the Conference in Honor of
Raymond Stora's 80th Birthday, LAPTh, Annecy, July 8, 2011}
\end{center}

\end{titlepage}

\baselineskip 6 mm



\section{From Extremal Black Holes...}

Black holes, one of the most stunning consequences of General
Relativity, enjoy thermodynamical properties in a generalized phase
space whose quantum mechanical attributes that are their
Arnowitt-Deser-Misner (ADM) mass \cite{ADM}, charge, spin and scalar
charges (see for instance
\cite{Ferrara:2008hwa,SUSY,FGK,reviewSF,reviewK}). They can be
regarded as probes of the quantum regime of any fundamental theory
of gravity and, as such, they are naturally investigated within the
framework of superstring and M-theory. Unlike Schwarzschild black
holes, charged (Reissner-Nordstr\"{o}m) and/or spin (Kerr-Newman)
black holes can be \textit{extremal}, \textit{i.e.} with vanishing
temperature for non-zero entropy, in which case their event and
Cauchy horizons coincide. In formulae, the extremality parameter is
given by
\begin{equation}
c=2ST=\frac{1}{2}(r_{+}-r_{-})\rightarrow 0\,,
\end{equation}
where $c$ measures the surface gravity and $S=log\ \mathcal{N}$ is the black
hole entropy which counts the number $\mathcal{N}$ of microstates. In
(semi)classical (super)gravity, $S$ is given by the celebrated
Bekenstein-Hawking area-entropy formula \cite{BH}
\begin{equation}
S=\frac{1}{2}A_{H}=\pi R_{H}^{2}\,=\pi V_{BH,crit}\left( Q\right) ,
\end{equation}
where $R_{H}$ is the effective radius of the horizon, and $V_{BH}\left( \phi
,Q\right) $ is the so-called black hole effective potential (a function of
the scalar fields $\phi $ and of the electric and magnetic charges $Q$; see
below), defined within the \textit{Attractor Mechanism} \cite{SUSY,FGK} :
its critical points in the scalar manifold correspond to attractive scalar
trajectories towards the horizon itself. For extremal charged black holes, $%
R_{H}$ must respect the symmetries of the theory and in particular it must
depend only on the electric and magnetic charges and not on the scalar field
values \cite{GZreview}. Therefore, also the entropy will depend only on the
charges and it will take particular expressions depending on the duality
symmetries of the given model.

Another important feature of extremal black holes is that their horizon
geometry of spacetime is universal, and in four dimensions it is given by
the $AdS_{2}\times S_{2}$ Bertotti-Robinson metric. Actually, extremal black
holes behave as solitons interpolating between maximally symmetric
geometries of (super)spacetime: Minkowski at spacial infinity and the
conformally flat near-horizon metric \cite{Duff:1994an,GibbTown}.

In a supergravity, it should be recalled that, remarkably, the common radius
of $AdS_{2}\times S_{2}$, and therefore the entropy, can be actually
computed using the underlying non-compact electric-magnetic duality \cite{GZreview}%
.

In the last few years it has become clear that the scalar field dynamics for
the extremal black holes can be entirely encoded into a (charge orbit dependent) real \textit{%
``superpotential''} function $W(\phi ,Q)$ whose critical points
coincide with a whole class of critical points of $V_{BH}$ itself.
For supersymmetric $N=2$ flows $W=\left| Z\right| $, where $Z$ is
the
central extension of the local supersymmetry algebra; for $N>2$ $\frac{1}{N}$%
-BPS flows, $\left| Z\right| ^{2}$ is replaced by the highest
eigenvalues of $ZZ^{{\dag }}$, where $Z\equiv Z_{AB}$ is the central
charge matrix \cite {FM-MS-1}.

Remarkably, such a function $W$ can be shown to exist also for non
supersymmetric configurations, in which case it is called the
\textit{``fake superpotential''} \cite{Ceresole:2007wx} because of
the similarity with the setup of \textit{``fake supergravities'',}
and applications in domain-wall physics \cite{fake}. When the
attractors are regular, the $W$ function has a minimum for $\phi
^{i}=\phi _{H}^{i}$, and its horizon value gives the entropy of the
configuration
\begin{equation}
S=\frac{1}{4}A_{H}=\pi W_{H}^{2}(Q)=\pi W_{\mathrm{crit}}^{2}(\phi
_{h}^{i}(Q),Q)
\end{equation}
according to the aforementioned Bekenstein-Hawking formula. However, if $%
W$ has a runaway behavior in moduli space, $\phi _{H}\rightarrow \infty
\,W\rightarrow 0$ (which is not acceptable in $D=4$), the corresponding
black hole solutions are singular. Then the scalar fields are never
stabilized within the boundaries of moduli space, there are no attractors
and the entropy of the extremal configuration vanishes.

In order to describe a static, spherically symmetric, asymptotically
flat, extremal dyonic black hole background in the extremal case,
$c=2ST=0$, the metric \textit{Ansatz} reads \cite{FGK}
\begin{equation}
ds^{2}=-e^{2U}dt^{2}+e^{-2U}\left[ \frac{d\tau ^{2}}{\tau ^{4}}+\frac{1}{%
\tau ^{2}}(d\theta ^{2}+\sin ^{2}{\theta }d\phi ^{2})\right] \ .
\label{s-t-metric}
\end{equation}
with the field strength $F_{\mu \nu }^{\Lambda }$ for $n_{V}$ vectors ($%
\Lambda =1,\ldots ,n_{V}$) and its dual $G_{\Lambda \mu \nu }=\frac{\delta
\mathcal{L}}{\delta F_{\mu \nu }^{\Lambda }}$ given by
\begin{eqnarray}
F &=&e^{2U}C\mathcal{M}(\phi ^{i})Qdt\wedge d\tau +Q\sin {\theta }d\theta
\wedge d\phi \, \\
F &=&\left(
\begin{array}{c}
F_{\mu \nu }^{\Lambda } \\
G_{\Lambda \mu \nu }
\end{array}
\right) \frac{dx^{\mu }dx^{\nu }}{2}\,.
\end{eqnarray}
The electric and magnetic charge vector $Q\equiv \left( p^{\Lambda
},q_{\Lambda }\right) ^{T}$ is defined as
\begin{equation}
q_{\Lambda }\equiv \frac{1}{4\pi }\int_{S_{\infty }^{2}}G_{\Lambda
}\,,\qquad p^{\Lambda }\equiv \frac{1}{4\pi }\int_{S_{\infty
}^{2}}F_{\Lambda }\,.\label{Q}
\end{equation}
\noindent $\mathcal{M}(\phi )$ is a $2n_{V}\times 2n_{V}$ real, symmetric,
negative-definite $Sp(2n_{V},\Bbb{R})$ matrix, satisfying $\mathcal{M}\Bbb{C}%
\mathcal{M}=\Bbb{C}$ ($\Bbb{C}$ denoting the symplectic metric), and given
by
\begin{equation}
\mathcal{M}(\phi )=\left(
\begin{array}{cc}
I+RI^{-1}R & -RI^{-1} \\
-I^{-1}R & I^{-1}
\end{array}
\,.\right)
\end{equation}
where $I\equiv $Im$\,\mathcal{N}_{\Lambda \Sigma }$, $R\equiv $Re$\,\mathcal{%
N}_{\Lambda \Sigma }$, with $\mathcal{N}_{\Lambda \Sigma }$ denoting the
scalar-dependent vector kinetic matrix appearing in the D=4 lagrangian
density of the Maxwell-Einstein-scalar system ($i=1,...,n_{V}-1$)
\begin{equation}
\mathcal{L}=-\frac{R}{2}+\frac{1}{2}g_{ij}(\phi )\partial _{\mu }\phi
^{i}\partial ^{\mu }\phi ^{j}+I_{\Lambda \Sigma }F^{\Lambda }\wedge ^{\ast
}F^{\Sigma }+R_{\Lambda \Sigma }F^{\Lambda }\wedge F^{\Sigma }\,.  \label{L}
\end{equation}

The aforementioned black hole effective potential $V_{BH}$
\cite{SUSY}, governing the radial evolution of the scalar fields in
the black hole background (\ref{s-t-metric}) enjoys a very simple
expression in terms of the matrix $\mathcal{M}$, namely:
\begin{equation}
V_{BH}=-\frac{1}{2}Q^{T}\mathcal{M}Q\,.  \label{eff}
\end{equation}
As pioneered in \cite{FGK}, such a function arises upon reducing the
general $D=4$ Lagrangian (\ref{L}) in the background
(\ref{s-t-metric}) to the $D=1$ \textit{almost geodesic} action
describing the radial evolution of the $n_{V}$ scalar fields $\{
U(\tau ),\phi ^{i}(\tau )\} $:
\begin{equation}
S_{D=1}=\int \left( U^{\prime }+g_{ij}\phi ^{\prime i}\phi ^{\prime
j}+e^{2U}V_{BH}\left( \phi \left( \tau \right) ,p,q\right) \right)
d\tau ,
\end{equation}
where $\tau $ is the $D=1$ \textit{affine} evolution parameter in
the extremal black hole background (\ref{s-t-metric}), and prime
stands for differentiation with respect to it. In order to have the
same equations of motion of the original theory, the action must be
complemented with the Hamiltonian constraint (in the extremal case)
\cite{FGK}
\begin{equation}
\left( U^{\prime }\right) ^{2}+g_{ij}\phi ^{\prime i}\phi ^{\prime
j}-e^{2U}V_{BH}\left( \phi \left( \tau \right) ,p,q\right) =0.
\end{equation}
The black hole effective potential can be written in terms of the
superpotential $W$ as
\begin{equation}
V_{BH}=W^{2}+2g^{i{j}}\partial _{i}W\partial _{j}W\,\,.  \label{VW}
\end{equation}
This formula can be viewed as a differential equation defining $W$
for a given black hole effective potential $V_{BH}$, and it can lead
to multiple choices: only one of those will corresponds to BPS
solutions (i.e. to the usual superpotential), while others will be
associated to non-BPS ones. In both cases, $W$ allows to rewrite the
ordinary \textit{second-order} supergravity equations of motion
\begin{eqnarray}
\frac{d^{2}U}{d\tau ^{2}} &=&e^{2U}V_{BH} \\
\frac{d^{2}\phi ^{i}}{d\tau ^{2}} &=&g^{i\bar{\jmath}}\frac{\partial V_{BH}}{%
\partial \phi _{\bar{\jmath}}}e^{2U}\,,
\end{eqnarray}
as \textit{first-order }flow equations, defining the radial
evolution of the scalar fields $\phi ^{i}$ and the warp factor $U$
from asymptotic infinity towards the black hole horizon
\cite{Ceresole:2007wx} :
\begin{equation}
U^{\prime }=-e^{U}W\,,\qquad \qquad \phi ^{\prime i}=-2e^{U}g^{i{j}}\partial
_{j}W\,.
\end{equation}

Beside the horizon entropy $S_{BH}=\pi W_{H}^{2}$ and the first order flows,
the value at radial infinity of the superpotential $W$ also encodes other
basic property of the extremal black hole, which are its $ADM$ mass \cite
{ADM}, given by
\begin{equation}
M_{ADM}(\phi _{0}^{i},Q)\equiv W(\phi _{\infty },Q),
\end{equation}
and the \textit{scalar charges}
\begin{equation}
\Sigma ^{i}\equiv \phi _{\infty }^{\prime i}=-2g^{ij}(\phi _{\infty })\frac{%
\partial W}{\partial \phi ^{i}}(\phi _{\infty },Q).
\end{equation}

For $N\geq 2$, the fake superpotential for the non-BPS branch\cite
{Dabholkar:2006tb,larsen} has been computed for wide classes of models \cite
{Ceresole:2007wx,Cardoso5d,Andrianopoli:2007gt,Bellucci:2008sv,Ceresole:2009iy,universality, Bossard:2009we, varinonBPS,Andrianopoli:2009je}%
, based on \textit{symmetric} geometries of moduli spaces, using as
a tool the U-duality symmetry of the underlying supergravity. A
universal procedure for its construction in $N=2$ special geometries
has been established \cite {Ceresole:2009iy,universality}, which
generalizes the results obtained for the so-called $N=2$ $STU$ model
\cite{stu}.

\section{...to Groups of Type $E_{7}$}

As yielded by the treatment above, the black hole entropy $S$ is
invariant under the electric-magnetic duality, within the framework
firstly defined in \cite{GZreview}, in which the non-compact
U-duality group has a symplectic action both on the charge vector
$Q$ (\ref{Q}) and on the scalar fields (through the definition of a
\textit{flat} symplectic bundle \cite {Strominger-SKG} over the
scalar manifold itself); see \textit{e.g.} \cite{dW-review} for a
review. The latter property makes relevant the mathematical notion
of groups of type $E_{7}$.

The first axiomatic characterization of groups of type $E_{7}$
 through a module (irrep.) was given in 1967 by Brown \cite{Brown-Groups-of-type-E7}%
. A group $G$ of type $E_{7}$ is a Lie group endowed with a
representation $\mathbf{R}$ such that:

\begin{enumerate}
\item  $\mathbf{R}$ is \textit{symplectic}, \textit{i.e.} :
\begin{equation}
\exists !\mathbb{C}_{\left[ MN\right] }\equiv \mathbf{1\in R\times }_{a}\mathbf{%
R;}\label{sympl-metric}
\end{equation}
(the subscripts ``$s$'' and ``$a$''  stand for symmetric and
skew-symmetric throughout) in turn, $\mathbb{C}_{\left[ MN\right] }$
defines a non-degenerate skew-symmetric bilinear form
(\textit{symplectic product}); given two different charge vectors
$Q_{1}$ and $Q_{2}$ in $\mathbf{R}$, such a bilinear form is defined
as
\begin{equation}
\left\langle Q_{1},Q_{2}\right\rangle \equiv Q_{1}^{M}Q_{2}^{N}\mathbb{C}%
_{MN}=-\left\langle Q_{2},Q_{1}\right\rangle ;
\end{equation}

\item  $\mathbf{R}$ admits a unique rank-$4$ completely symmetric primitive $%
G$-invariant structure, usually named $K$-tensor
\begin{equation}
\exists !\mathbb{K}_{\left( MNPQ\right) }\equiv \mathbf{1\in }\left[ \mathbf{%
R\times R\times R\times R}\right] _{s}\mathbf{;}
\end{equation}
thus, by contracting the $K$-tensor with the same charge vector $Q$ in $%
\mathbf{R}$, one can construct a rank-4 homogeneous $G$-invariant
polynomial, named $\mathcal{I}_{4}$:
\begin{equation}
\mathcal{I}_{4}\left( Q\right) \equiv \frac{1}{2}\mathbb{K}%
_{MNPQ}Q^{M}Q^{N}Q^{P}Q^{Q},\label{I4}
\end{equation}
which corresponds to the evaluation of the rank-$4$ symmetric form $\mathbf{q%
}$ induced by the $K$-tensor on four identical modules $\mathbf{R}$:
\begin{equation}
\mathcal{I}_{4}\left( Q\right) =\frac{1}{2}\left. \mathbf{q}\left(
Q_{1},Q_{2},Q_{3},Q_{4}\right) \right| _{Q_{1}=Q_{2}=Q_{3}=Q_{4}\equiv
Q}\equiv \frac{1}{2}\left[ \mathbb{K}_{MNPQ}Q_{1}^{M}Q_{2}^{N}Q_{3}^{P}Q_{4}^{Q}%
\right] _{Q_{1}=Q_{2}=Q_{3}=Q_{4}\equiv Q}.
\end{equation}
A famous example of \textit{quartic} invariant in $G=E_{7}$ is the \textit{%
Cartan-Cremmer-Julia} invariant \cite{Cartan}, constructed out of the
fundamental irrep. $\mathbf{R}=\mathbf{56}$.

\item  if a trilinear map $T\mathbf{:R\times R\times R}\rightarrow \mathbf{R}
$ is defined such that
\begin{equation}
\left\langle T\left( Q_{1},Q_{2},Q_{3}\right) ,Q_{4}\right\rangle =\mathbf{q}%
\left( Q_{1},Q_{2},Q_{3},Q_{4}\right) ,
\end{equation}
then it holds that
\begin{equation}
\left\langle T\left( Q_{1},Q_{1},Q_{2}\right) ,T\left(
Q_{2},Q_{2},Q_{2}\right) \right\rangle =\left\langle
Q_{1},Q_{2}\right\rangle \mathbf{q}\left( Q_{1},Q_{2},Q_{2},Q_{2}\right) .
\end{equation}
This last property makes the group of type $E_{7}$ amenable to a
treatment in terms of (rank-3) Jordan algebras and related
Freudenthal triple systems.
\end{enumerate}

Remarkably, groups of type $E_{7}$, appearing in $D=4$ supergravity as $U$%
-duality groups, admit a $D=5$ uplift to groups of type $E_{6}$, as
well as a $D=3$ downlift to groups of type $E_{8}$; see
\cite{Truini}. It should also be recalled that split form of
exceptional Lie groups appear in the
exceptional Cremmer-Julia \cite{CJ} sequence $E_{D\left( D\right) }$ of $U$%
-duality groups of $M$-theory compactified on a $D$-dimensional torus, in $%
D=3,4,5$.

It is intriguing to notice that the first paper on groups of type
$E_{7}$ was written about a decade before the discovery of of
extended ($N=2$) supergravity \cite{FVN}, in which electromagnetic
duality symmetry was observed \cite{FSZ}. The connection of groups
of type $E_{7}$
to supergravity can be summarized by stating that all $2\leq N\leq 8$%
-extended supergravities in $D=4$ with symmetric scalar manifolds ${\frac{G}{%
H}}$ have $G$ of type $E_{7}$ \cite {Borsten:2009zy,Ferrara:2011gv},
with the exception of $N=2$ group $G=U(1,n)$ and $N=3$ group
$G=U(3,n)$. These latter in fact have a quadratic invariant
Hermitian form $\left( Q_{1},\overline{Q}_{2}\right) $, whose
imaginary part is the symplectic (skew-symmetric) product and whose
real part is the symmetric quadratic invariant
$\mathcal{I}_{2}\left( Q\right) $ defined as follows
\begin{eqnarray}
\mathcal{I}_{2}\left( Q\right)  &\equiv &\left[ \text{Re}\left( Q_{1},%
\overline{Q}_{2}\right) \right] _{Q_{1}=Q_{2}}; \\
\left\langle Q_{1},\overline{Q}_{2}\right\rangle  &=&-\text{Im}\left( Q_{1},%
\overline{Q}_{2}\right) .
\end{eqnarray}
Thus, the fundamental representations of pseudo-unitary groups
$U(p,n)$, which have a Hermitian quadratic invariant form, do not
strictly qualify for groups of type $E_{7}$.

In theories with groups of type $E_{7}$, the Bekenstein-Hawking
black hole entropy is given by
\begin{equation}
S=\pi \sqrt{\left| \mathcal{I}_{4}\left( Q\right) \right| },\label{S-I4}
\end{equation}
as it was proved for the case of $G=E_{7(7)}$ (corresponding to $N=8$
supergravity) in \cite{Kallosh:1996uy}. For $N=2$ group $G=U(1,n)$ and $N=3$
group $G=U(3,n)$ the analogue of (\ref{S-I4}) reads
\begin{equation}
S=\pi \left| \mathcal{I}_{2}\left( Q\right) \right| .
\end{equation}

For $3 < N\leq 8$ the following groups of type $E_{7}$ are relevant:
$E_{7\left( 7\right) }$, $SO^*(12)$, $SU(1,5)$,
$SL(2,\mathbb{R})\times {SO(6,n)}$; see Table 1. In $N=2$ cases of
\textit{symmetric} vector multiplets' scalar manifolds, there are
$6$ groups of type $E_{7}$ \cite{Gunaydin:1983rk} : $E_{7\left(
-25\right) }$, $SO^*(12)$, $SU(3,3)$, $Sp(6,\mathbb{R})$,
$SL(2,\mathbb{R})$, and $SL(2,\mathbb{R})\times SO(2,n)$; see Table
2. Here $n$ is the integer describing the number of matter (vector)
multiplets for $N=4,3,2$.

\begin{table}[t]
\begin{center}
\begin{tabular}{|c||c|c|}
\hline $N$& $
\begin{array}{c}
\\
$G$ \\
~
\end{array}
$ & $
\begin{array}{c}
\\
  \mathbf{ R}
\\
~
\end{array}

$ \\ \hline\hline $
\begin{array}{c}
\\
3 \\
~
\end{array}
$ & $U(3,n)$ & $ \mathbf{(3+n)}$     \\ \hline $
\begin{array}{c}
\\
4 \\
~
\end{array}
$ & $SL(2, \mathbb{R})\times {SO(6,n)}$ & $\mathbf{(2, 6+n)}$   \\
\hline $
\begin{array}{c}
\\
5 \\
~
\end{array}
$ & $SU(1,5)$ & $ \mathbf{ 20}$   \\ \hline $
\begin{array}{c}
\\
6 \\
~
\end{array}
$ & $SO^*(12)$ & $\mathbf{ 32}$
 \\ \hline
$
\begin{array}{c}
\\
8 \\
~
\end{array}
$ & $E_{7\left( 7\right) }$ & $\mathbf{ 56}$   \\ \hline
\end{tabular}
\end{center}
\caption{ $N\geq 3$ supergravity sequence of groups $G$ of  the
corresponding ${G\over H}$ symmetric spaces, and their symplectic
representations  $\mathbf{R}$}
\end{table}

\begin{table}[t]
\begin{center}
\begin{tabular}{|c||c|}
\hline $
\begin{array}{c}
\\
$G$\\
~
\end{array}
$ & $\mathbb{\mathbb{}}
\begin{array}{c}
\\
  \mathbf{ R}\\
~
\end{array}
$ \\ \hline\hline $
\begin{array}{c}
\\
{U(1,n)}
\end{array}
$ & $
\begin{array}{c}

\mathbf{(1+n)}\\

\end{array}
$ \\ \hline $
\begin{array}{c}
\\
{SL(2, \mathbb{R})}\times SO(2,n) ~ ~

\end{array}
$ & $
\begin{array}{c}
\mathbf{(2, 2+n)}

\end{array}
$ \\ \hline $
\begin{array}{c}
\\
SL(2, \mathbb{R}) ~
\end{array}
$ & $
\begin{array}{c}
\\
\mathbf{4}
\end{array}
$ \\ \hline $
\begin{array}{c}
\\
Sp(6,\mathbb{R})~
\end{array}
$ & $
\begin{array}{c}
\\
\mathbf{14}' \\

\end{array}
$ \\ \hline $
\begin{array}{c}
\\
SU(3,3)\end{array} $ & $
\begin{array}{c}
\\
\mathbf{20} ~
\end{array}
$ \\ \hline $
\begin{array}{c}
\\
SO^{\ast }(12)~
\end{array}
$ & $
\begin{array}{c}
\\
\mathbf{32} ~
\end{array}
$ \\ \hline $
\begin{array}{c}
\\
E_{7\left( -25\right) } ~
\end{array}
$ & $
\begin{array}{c}
\\
\mathbf{56} ~
\end{array}
$ \\ \hline
\end{tabular}
\end{center}
\caption{$N=2$ choices of groups $G$ of the  ${G\over H}$ symmetric
spaces and their symplectic representations  $\mathbf{R}$. The last
four lines refer to ``magic" $N=2$ supergravities.}
\end{table}

\section{Orbits}

We here report some results on the stratification of the
$\mathbf{R}$ irrep. space of simple groups $G$ $E_{7}$. For a recent
account, with a detailed list of Refs., see \textit{e.g.}
\cite{Small-Orbits-Phys}.

In supergravity, this corresponds to $U$-duality invariant
constraints defining the charge orbits of a single-centered extremal
black hole, namely of the $G$-invariant conditions defining the
\textit{rank} of the dyonic charge vector $Q$ (\ref{Q}) in
$\mathbf{R}$ as an element of the corresponding Freudenthal triple
system (FTS) (see \cite{Ferrar,Krut}, and
Refs. therein). The symplectic indices $M=1,...,\mathbf{f}$ ($\mathbf{f}%
\equiv $dim$_{\mathbb{R}}\mathbf{R}\left( G\right) $) are raised and
lowered
with the symplectic metric $\mathbb{C}_{MN}$ defined by (\ref{sympl-metric}%
). By recalling the definition (\ref{I4}) of the unique primitive rank-4 $G$%
-invariant polynomial constructed with $Q$ in $\mathbf{R}$, the
\textit{rank} of a non-null $Q$ as an element of FTS$\left( G\right)
$ ranges from four to one, and it is manifestly $G$-invariantly
characterized as follows:

\begin{enumerate}
\item  \textit{rank}$\left( Q\right) =4$. This corresponds to ``large''
extremal black holes, with non-vanishing area of the event horizon
(exhibiting \textit{Attractor Mechanism} \cite{SUSY,FGK}):
\begin{equation}
\mathcal{I}_{4}\left( Q\right) < 0, ~ ~ or ~ ~
\mathcal{I}_{4}\left( Q\right) >0  \label{rank=4}
\end{equation}

\item  \textit{rank}$\left( Q\right) =3$. This corresponds to ``small''
\textit{lightlike} extremal black holes, with vanishing area of the
event horizon:
\begin{equation}
\begin{array}{l}
\mathcal{I}_{4}\left( Q\right) =0; \\
\\
T\left( Q,Q,Q\right) \neq 0.
\end{array}
\end{equation}

\item  \textit{rank}$\left( Q\right) =2$. This corresponds to ``small''
\textit{critical} extremal black holes:
\begin{equation}
\begin{array}{l}
T\left( Q,Q,Q\right) =0; \\
\\
3T\left( Q,Q,P\right) +\left\langle Q,P\right\rangle Q\neq 0.
\end{array}
\end{equation}

\item  \textit{rank}$\left( Q\right) =1$. This corresponds to ``small''
\textit{doubly-critical} extremal BHs
\cite{FG,Ferrara-Maldacena,DFL-0-brane}:
\begin{equation}
3T\left( Q,Q,P\right) +\left\langle Q,P\right\rangle Q=0,~\forall
P\in \mathbf{R}.\label{rank=1}
\end{equation}
\end{enumerate}

Let us consider the doubly-criticality condition (\ref{rank=1}) more
in detail. \textit{At least} for \textit{simple} groups of type
$E_{7}$, the following holds:
\begin{eqnarray}
\mathbf{R}\times _{s}\mathbf{R} &=&\mathbf{Adj}+\mathbf{S};  \label{symm-1} \\
\mathbf{R}\times _{a}\mathbf{R} &=&\mathbf{1}+\mathbf{A},
\label{skew-symm}
\end{eqnarray}
where $\mathbf{S}$ and $\mathbf{A}$ are suitable irreps.. For example, for $%
G=E_{7}$ ($\mathbf{R}=\mathbf{56}$, $\mathbf{Adj}=\mathbf{133}$) one
gets (see \textit{e.g.} \cite{Slansky})
\begin{eqnarray}
\left( \mathbf{56}\times \mathbf{56}\right) _{s} &=&\mathbf{133}+\mathbf{1463%
}; \\
\left( \mathbf{56}\times \mathbf{56}\right) _{a}
&=&\mathbf{1}+\mathbf{1539}.
\end{eqnarray}
For such groups, one can construct the projection operator on $\mathbf{Adj}%
\left( G\right) $:
\begin{eqnarray}
\mathcal{P}_{AB}^{~~CD} &=&\mathcal{P}_{\left( AB\right) }^{~~\left(
CD\right) }; \\
\mathcal{P}_{AB}^{~~CD}\frac{\partial ^{2}\mathcal{I}_{4}}{\partial
Q^{C}\partial Q^{D}} &=&\left. \frac{\partial
^{2}\mathcal{I}_{4}}{\partial
Q^{A}\partial Q^{B}}\right| _{\mathbf{Adj}\left( G\right) }; \\
\mathcal{P}_{AB}^{~~CD}\mathcal{P}_{CD}^{~~EF}\frac{\partial ^{2}\mathcal{I}%
_{4}}{\partial Q^{E}\partial Q^{F}}
&=&\mathcal{P}_{AB}^{~~EF}\frac{\partial
^{2}\mathcal{I}_{4}}{\partial Q^{E}\partial Q^{F}},
\end{eqnarray}
where (recall (\ref{symm-1}))
\begin{eqnarray}
\frac{\partial ^{2}\mathcal{I}_{4}}{\partial Q^{A}\partial Q^{B}}
&=&\left.
\frac{\partial ^{2}\mathcal{I}_{4}}{\partial Q^{A}\partial Q^{B}}\right| _{%
\mathbf{Adj}\left( G\right) }+\left. \frac{\partial ^{2}\mathcal{I}_{4}}{%
\partial Q^{A}\partial Q^{B}}\right| _{\mathbf{S}\left( G\right) }; \\
\left. \frac{\partial ^{2}\mathcal{I}_{4}}{\partial Q^{A}\partial Q^{B}}%
\right| _{\mathbf{Adj}\left( G\right) } &=&2\left( 1-\tau \right)
\left(
3\mathbb{K}_{ABCD}+\mathbb{C}_{AC}\mathbb{C}_{BD}\right) Q^{C}Q^{D}; \\
\left. \frac{\partial ^{2}\mathcal{I}_{4}}{\partial Q^{A}\partial Q^{B}}%
\right| _{\mathbf{S}\left( G\right) } &=&2\left[ 3\tau \mathbb{K}%
_{ABCD}+\left( \tau -1\right) \mathbb{C}_{AC}\mathbb{C}_{BD}\right]
Q^{C}Q^{D},
\end{eqnarray}
where $\tau \equiv 2\mathbf{d/}\left[ \mathbf{f}\left( \mathbf{f}+1\right) %
\right] $, $\mathbf{d}\equiv $dim$_{\mathbb{R}}\left(
\mathbf{Adj}\left( G\right) \right) $. The explicit expression of
$\mathcal{P}_{AB}^{~~CD}$
reads\footnote{%
For related results in terms of a map formulated in the ``$4D/5D$
special coordinates'' symplectic frame (and thus manifestly
covariant under the $d=5$
$U$-duality group $G_{5}$), see \textit{e.g.} \cite{Shukuzawa,Yokota}.} ($%
\alpha =1,...,\mathbf{d}$):
\begin{equation}
\mathcal{P}_{AB}^{~~CD}=\tau \left( 3\mathbb{C}^{CE}\mathbb{C}^{DF}\mathbb{K}%
_{EFAB}+\delta _{(A}^{C}\delta _{B)}^{D}\right) =-t^{\alpha \mid
CD}t_{\alpha \mid AB},  \label{P-Adj}
\end{equation}
where the relation \cite{Exc-Reds} (see also \cite{Og-1})
\begin{equation}
\mathbb{K}_{MNPQ}=-\frac{1}{3\tau }t_{(MN}^{\alpha }t_{\alpha \mid PQ)}=-%
\frac{1}{3\tau }\left[ t_{MN}^{\alpha }t_{\alpha \mid PQ}-\tau \mathbb{C}%
_{M(P}\mathbb{C}_{Q)N}\right] ,  \label{rel-2}
\end{equation}
where
\begin{equation}
t_{MN}^{\alpha }=t_{\left( MN\right) }^{\alpha };~~t_{MN}^{\alpha }\mathbb{C}%
^{MN}=0
\end{equation}
is the symplectic representation of the generators of the Lie algebra $\frak{%
g}$ of $G$. Notice that $\tau <1$ is nothing but the ratio of the
dimensions of the adjoint $\mathbf{Adj}$ and rank-$2$ symmetric
$\mathbf{R}\times _{s}\mathbf{R}$ (\ref{symm-1}) reps. of $G$, or
equivalently the ratio of upper and lower indices of $t_{MN}^{\alpha
}$'s themselves.

\section{Special K\"{a}hler Geometry
\\and\textit{``Generalization''} of
Groups of Type $E_{7}$}

Here we would like to discuss the characterization of special
K\"{a}hler geometry (SKG) in terms of a suitable
\textit{``generalization'' }of the groups of type $E_{7}$, recently
proposed in \cite {T-Tensors} (for some preliminary discussion, see
also Sec. 4 of \cite {FMY-FD-1}).

As obtained in \cite{CFMZ1} (see Eq. (5.36) therein), the following
real function, which we dub \textit{``entropy functional''}, can be
defined on
the vector multiplets' scalar manifold\footnote{%
Note that the expression (\ref{I4-N=2-symm}) is independent on the
choice of
the symplectic frame and manifestly invariant under diffeomorphisms in $%
\mathbf{M}$.} $\mathbf{M}$:
\begin{equation}
\mathbb{I}_{4}=\left( \left| Z\right| ^{2}-Z_{i}\overline{Z}^{i}\right) ^{2}+%
\frac{2}{3}i\left( Z\overline{C}_{\overline{i}\overline{j}\overline{k}}Z^{%
\overline{i}}Z^{\overline{j}}Z^{\overline{k}}-\overline{Z}C_{ijk}\overline{Z}%
^{i}\overline{Z}^{j}\overline{Z}^{k}\right) -g^{i\overline{i}}C_{ijk}%
\overline{C}_{\overline{i}\overline{l}\overline{m}}\overline{Z}^{j}\overline{%
Z}^{k}Z^{\overline{l}}Z^{\overline{m}}.  \label{I4-N=2-symm}
\end{equation}
$Z_{i}\equiv D_{i}Z$ are the so-called ``matter charges'' ($%
D_{i}$ stands for the K\"{a}hler-covariant differential operator;
see \textit{e.g.} \cite{CDF-rev} and \cite{N=2-Big} for notation and
further elucidation):
\begin{equation}
Z\equiv Q^{M}V^{N}\mathbb{C}_{MN};~Z_{i}\equiv Q%
^{M}V_{i}^{N}\mathbb{C}_{MN},  \label{Z-Zi-def}
\end{equation}
with $V^{M}$ denoting the vector of covariantly-holomorphic
symplectic sections of SKG, and $V_{i}^{M}\equiv D_{i}V^{M}$.
Furthermore, $C_{ijk}$ is the rank-$3$, completely symmetric,
covariantly holomorphic tensor of SKG (with K\"{a}hler weights
$\left( 2,-2\right) $) (see \textit{e.g.} \cite {CDF-1,CDF-2}):
\begin{equation}
\begin{array}{l}
C_{ijk}\equiv \mathbb{C}_{MN}\left( D_{i}D_{j}V^{M}\right) D_{k}V^{N}=-ig_{i%
\overline{l}}\overline{f}_{\Lambda
}^{\overline{l}}D_{j}D_{k}L^{\Lambda
}=D_{i}D_{j}D_{k}\mathcal{S}=e^{K}W_{ijk}; \\
\overline{f}_{\Lambda }^{\overline{l}}\left( \overline{D}\overline{L}_{%
\overline{s}}^{\Lambda }\right) \equiv \delta _{\overline{s}}^{\overline{l}%
},~\mathcal{S}\equiv -iL^{\Lambda }L^{\Sigma }\text{Im}\left(
F_{\Lambda
\Sigma }\right) ,~\overline{\partial }_{\overline{l}}W_{ijk}=0; \\
\overline{D}_{\overline{i}}C_{jkl}=0; \\
D_{[i}C_{j]kl}=0,
\end{array}
\label{C}
\end{equation}
the last property being a consequence, through the covariant
holomorphicity of $C_{ijk}$ and the SKG constraint on the Riemann
tensor (see \textit{e.g.} \cite{CDF-1,CDF-2,VP-what-SKG})
\begin{equation}
R_{j\overline{k}l\overline{m}}=-g_{j\overline{k}}g_{l\overline{m}}-g_{j%
\overline{m}}g_{l\overline{k}}+g^{i\overline{i}}C_{ijl}\overline{C}_{%
\overline{i}\overline{k}\overline{m}},  \label{SKG-constraint}
\end{equation}
of the Bianchi identities satisfied by the Riemann tensor $R_{i\overline{j}k%
\overline{l}}$.

Furthermore, $\mathbb{I}_{4}$ is an order-$4$ homogeneous polynomial
in the fluxes $\mathcal{Q}$; this allows for the definition of the
$Q$-independent rank-$4$ completely symmetric tensor $\Omega
_{MNPQ}$ \cite {FMY-FD-1}, whose general expression we explicitly
compute here:
\begin{eqnarray}
\Omega _{MNPQ} &\equiv &2\frac{\partial ^{4}\mathbb{I}_{4}}{\partial
Q^{(M}\partial Q^{N}\partial Q^{P}\partial
Q^{Q)}}  \label{Omega-tensor} \\
&=&2V_{(M}V_{N}\overline{V}_{P}\overline{V}_{Q)}+2V_{i\mid (M}\overline{V}%
_{N}^{i}V_{j\mid P}\overline{V}_{Q)}^{j}-4V_{(M}\overline{V}_{N}V_{i\mid P}%
\overline{V}_{Q)}^{i}  \notag \\
&&+\frac{2}{3}\left( V_{(M}V_{N}^{\overline{i}}V_{P}^{\overline{j}}\overline{%
D}_{\overline{i}}\overline{V}_{\overline{j}\mid Q}+\overline{V}_{(M}%
\overline{V}_{N}^{i}\overline{V}_{P}^{j}D_{i}V_{j\mid Q}\right)   \notag \\
&&-2g^{i\overline{i}}\overline{V}_{(M}^{j}V_{N}^{\overline{l}}D_{i}V_{j\mid
N}\overline{D}_{\overline{i}}\overline{V}_{\overline{l}\mid Q)},
\label{Omega-tensor-2}
\end{eqnarray}
where the SKG defining relation (see \textit{e.g.} \cite
{CDF-1,CDF-2,VP-what-SKG})
\begin{equation}
D_{i}D_{j}V^{M}\equiv D_{i}V_{j}^{M}=iC_{ijk}\overline{V}^{k\mid M}
\label{SKG-rel-1}
\end{equation}
has been used in order to recast (\ref{Omega-tensor}) in terms of $V^{M}$, $%
V_{i}^{M}$ and $D_{i}V_{j}^{M}$ only.

Some further elaborations are possible; \textit{e.g.}, by using
(\ref {SKG-constraint}), $\mathbb{I}_{4}$ (\ref{I4-N=2-symm}) and
$\Omega _{MNPQ}$ (\ref{Omega-tensor-2}) can respectively be
rewritten as
\begin{eqnarray}
\mathbb{I}_{4} &=&\left| Z\right| ^{4}-\left(
Z_{i}\overline{Z}^{i}\right)
^{2}-2\left| Z\right| ^{2}Z_{i}\overline{Z}^{i}+\frac{2}{3}i\left( Z%
\overline{C}_{\overline{i}\overline{j}\overline{k}}Z^{\overline{i}}Z^{%
\overline{j}}Z^{\overline{k}}-\overline{Z}C_{ijk}\overline{Z}^{i}\overline{Z}%
^{j}\overline{Z}^{k}\right) -\mathcal{R};  \label{I4-N=2-symm-2} \\
&&  \notag \\
\Omega _{MNPQ} &=&2V_{(M}V_{N}\overline{V}_{P}\overline{V}_{Q)}-2V_{i\mid (M}%
\overline{V}_{N}^{i}V_{j\mid P}\overline{V}_{Q)}^{j}-4V_{(M}\overline{V}%
_{N}V_{i\mid P}\overline{V}_{Q)}^{i}  \notag \\
&&+\frac{2}{3}\left( V_{(M}V_{N}^{\overline{i}}V_{P}^{\overline{j}}\overline{%
D}_{\overline{i}}\overline{V}_{\overline{j}\mid Q}+\overline{V}_{(M}%
\overline{V}_{N}^{i}\overline{V}_{P}^{j}D_{i}V_{j\mid Q}\right)
-R_{MNPQ}, \label{Omega-tensor-3}
\end{eqnarray}
where the \textit{sectional curvature of matter charges}
(\textit{cfr.} Eq. (5.3) of \cite{Raju-1}; also note that
(\ref{R-call-def}) is different from the definition given by Eq.
(3.1.1.2.11) of \cite{Kallosh-rev}))
\begin{equation}
\mathcal{R}\equiv R_{i\overline{j}k\overline{l}}\overline{Z}^{i}Z^{\overline{%
j}}\overline{Z}^{k}Z^{\overline{l}},  \label{R-call-def}
\end{equation}
and the corresponding rank-$4$ completely symmetric tensor
\begin{equation}
R_{MNPQ}\equiv \frac{\partial ^{4}\mathcal{R}}{\partial Q%
^{(M}\partial Q^{N}\partial Q^{P}\partial Q%
^{Q)}}=R_{i\overline{j}k\overline{l}}\overline{V}_{(M}^{i}V_{N}^{\overline{j}%
}\overline{V}_{P}^{k}V_{Q)}^{\overline{l}},  \label{R-sympl-tensor}
\end{equation}
have been introduced. Note that $R_{MNPQ}$ can be regarded as the
completely symmetric part of the \textit{``symplectic pull-back''}
(through the
symplectic sections $V_{i}^{M}$) of the Riemann tensor $R_{i\overline{j}k%
\overline{l}}$ of $\mathbf{M}$.

Thus, SKG can be associated to a \textit{generalization} of the
class of groups of type $E_{7}$, based on $\mathbb{I}_{4}$ and the
corresponding (generally field-dependent, non-constant) $\Omega
$-structure:
\begin{equation}
\text{SKG~}:\left\{
\begin{array}{l}
\Omega _{MNPQ}:D_{i}\Omega _{MNPQ}=\partial _{i}\Omega _{MNPQ}\neq 0; \\
\\
\mathbb{I}_{4}\equiv \frac{1}{2}\Omega _{MNPQ}Q^{M}Q^{N}%
Q^{P}Q^{Q}\Rightarrow D_{i}\mathbb{I}_{4}=\partial _{i}%
\mathbb{I}_{4}\neq 0.
\end{array}
\right.   \label{gen-SKG}
\end{equation}

\textit{Symmetric} K\"{a}hler spaces have a covariantly constant
Riemann tensor:
\begin{equation}
D_{i}R_{j\overline{k}l\overline{m}}=0.
\end{equation}
Within SKG, through the constraint (\ref{SKG-constraint}), this
implies the covariant constancy of the $C$-tensor (\ref{C}):
\begin{equation}
D_{(i}C_{j)kl}=D_{(i}C_{jkl)}=0,
\end{equation}
which in turn yields the relation:
\begin{equation}
C_{p(kl}C_{ij)n}g^{n\overline{n}}g^{p\overline{p}}\overline{C}_{\overline{n}%
\overline{p}\overline{m}}=\frac{4}{3}g_{\left( l\right| \overline{m}%
}C_{\left| ijk\right) }\Leftrightarrow g^{n\overline{n}}R_{\left(
i\right|
\overline{m}\left| j\right| \overline{n}}C_{n\left| kl\right) }=-\frac{2}{3}%
g_{\left( i\right| \overline{m}}C_{\left| jkl\right) }.
\label{symm}
\end{equation}
Equivalently, \textit{symmetric} SK manifolds can be characterized
by stating that their $\Omega _{MNPQ}$ is \textit{independent} of
the scalar fields themselves, and it matches the $\mathbb{K}$-tensor
$\mathbb{K}_{MNPQ}$
defining the rank-$4$ invariant $\mathbb{K}$-structure of the corresponding $%
U$-duality group of type $E_{7}$ \cite{Brown-Groups-of-type-E7} (see
also \textit{e.g.} \cite{Exc-Reds}, and Refs. therein).
Consequently, the corresponding \textit{``entropy functional''}
$\mathbb{I}_{4}$ (\ref {I4-N=2-symm}) is \textit{independent} of the
scalar fields themselves, and it is thus a \textit{constant}
function in $\mathbf{M}$, given by the unique
algebraically-independent $1$-centered $U$-duality invariant polynomial $%
I_{4}$:
\begin{equation}
\underset{(U\text{-duality~group~}G\text{ is~of~type~}E_{7})}{\text{\textit{%
symmetric}~SKG}}\Rightarrow \left\{
\begin{array}{l}
\Omega _{MNPQ}=\mathbb{K}_{MNPQ}\Rightarrow D_{i}\Omega
_{MNPQ}=\partial
_{i}\Omega _{MNPQ}=0; \\
\\
\mathbb{I}_{4}=\mathcal{I}_{4}\equiv \frac{1}{2}\mathbb{K}_{MNPQ}Q%
^{M}Q^{N}Q^{P}Q^{Q}\Rightarrow D_{i}\mathbb{I}%
_{4}=\partial _{i}\mathbb{I}_{4}=0.
\end{array}
\right.   \label{symm-SKG}
\end{equation}
In turn, within \textit{symmetric} SKG, the pseudo-unitary
$U$-duality group $U\left( 1,s\right) $ (corresponding to $N=2$
\textit{minimally coupled} Maxwell-Einstein theory
\cite{Luciani,Gnecchi-1}) is
``degenerate'', in the aforementioned sense that the corresponding $\mathcal{%
I}_{4}$ actually is the square of the order-$2$ $U\left( 1,s\right) $%
-invariant polynomial $\mathcal{I}_{2}$. Indeed, $N=2$
\textit{minimally coupled} supergravity is characterized by
$C_{ijk}=0$, which plugged into (\ref{I4-N=2-symm}) (by taking (\ref
{symm-SKG}) into account) yields:
\begin{equation}
\underset{G=U\left( 1,s\right) }{\text{\textit{symmetric}~SKG~}}%
\Rightarrow \left\{
\begin{array}{l}
\Omega _{MNPQ}=\mathbb{K}_{MNPQ}\Rightarrow D_{i}\Omega
_{MNPQ}=\partial
_{i}\Omega _{MNPQ}=0; \\
C_{ijk}=0; \\
\mathbb{I}_{4}=\mathcal{I}_{4}=\left( \left| Z\right| ^{2}-Z_{i}\overline{Z}%
^{i}\right) ^{2}=\frac{1}{4}\mathcal{I}_{2}^{2}\Rightarrow D_{i}\mathbb{I}%
_{4}=\partial _{i}\mathbb{I}_{4}=0,
\end{array}
\right.
\end{equation}
where the normalization of \cite{MS-FMO-1} (see Eq. (2.15) therein)
has been adopted.

We conclude by recalling that, as noticed in \cite{CFMZ1} and in
\cite {FMY-FD-1}, the \textit{``entropic functional''
}$\mathbb{I}_{4}$ (\ref
{I4-N=2-symm}) is related to the \textit{geodesic potential} defined in the $%
D=4\rightarrow 3$ dimensional reduction of the considered $N=2$
theory. Under such a reduction, the $D=4$ vector multiplets' SK manifold $%
\mathbf{M}$ (dim$_{\mathbb{C}}=n_{V}$) enlarges to a
\textit{special} quaternionic K\"{a}hler manifold $\frak{M}$
(dim$_{\mathbb{H}}=n_{V}+1$)
given by $c$-map \cite{CFG,Ferrara-Sabharwal} of $\mathbf{M}$ itself : $%
\frak{M}=c\left( \mathbf{M}\right) $. By specifying Eq.
(\ref{I4-N=2-symm})
in the ``$4D/5D$ \textit{special coordinates' ''} symplectic frame, $\mathbb{%
I}_{4}$ matches the opposite of the function $h$ defined by Eq.
(4.42) of
\cite{dWVVP}, within the analysis of \textit{special} quaternionic K\"{a}%
hler geometry. This relation can be strengthened by observing that
the tensor $\Omega _{MNPQ}$ given by
(\ref{Omega-tensor})-(\ref{Omega-tensor-2}) is proportional to the
$\Omega $-tensor of quaternionic geometry, related to the
quaternionic Riemann tensor by Eq. (15) of \cite{Bagger-Witten}; for
further comments, see \cite{FMY-FD-1}.

\section*{Acknowledgments}

The reported results were obtained in collaboration with Laura
Andrianopoli, Stefano Bellucci, Leron Borsten, Bianca Letizia
Cerchiai, Anna Ceresole, Riccardo D'Auria, Gianguido Dall'Agata,
Mike Duff, Murat G\"{u}naydin, Renata Kallosh, Emanuele Orazi,
William Rubens, Raymond Stora, Mario Trigiante, Armen Yeranyan and
Bruno Zumino, which we gratefully acknowledge.

The work of S.F. is supported by the ERC Advanced Grant no. 226455,
\textit{``Supersymmetry, Quantum Gravity and Gauge Fields''} (\textit{%
SUPERFIELDS}), and in part by DOE Grant DE-FG03-91ER40662.

\end{document}